\documentclass[iop, twocolumn]{emulateapj}
\usepackage{apjfonts}
\usepackage{hyperref}

\newcommand\msun{{\mathrm{\,M_\odot}}}
\newcommand\grad{\mathrm{g_{rad}}}
\newcommand\teff{\mathrm{T_{\rm eff}}}
\newcommand\amlt{\alpha_{\rm MLT}}
\newcommand\logg{\log(g)}
\newcommand\Ysq{\mathrm{Y^2}}
\newcommand\minit{\mathrm{M_{init}}}

\shorttitle{Atomic Diffusion and Stellar Ages}
\shortauthors{Dotter et al.}

\journalinfo{Accepted for publication in ApJ}

\begin{document}

\title{The influence of atomic diffusion on stellar ages and chemical tagging}

\author{Aaron Dotter, Charlie Conroy, Phillip Cargile}
\affiliation{Harvard-Smithsonian Center for Astrophysics, Cambridge, MA 02138, USA}
\email{aaron.dotter@gmail.com}
\author{Martin Asplund}
\affiliation{Research School of Astronomy and Astrophysics, Australian National University, Canberra, ACT, Australia}

\begin{abstract}
In the era of large stellar spectroscopic surveys, there is emphasis on deriving not only stellar abundances but also ages for millions of stars. In the context of Galactic archeology, stellar ages provide a direct probe of the formation history of the Galaxy. We use the stellar evolution code MESA to compute models with atomic diffusion---with and without radiative acceleration---and extra mixing in the surface layers. The extra mixing consists of both density-dependent turbulent mixing and envelope overshoot mixing. Based on these models we argue that it is important to distinguish between initial, bulk abundances (parameters) and current, surface abundances (variables) in the analysis of individual stellar ages.  In stars that maintain radiative regions on evolutionary timescales, atomic diffusion modifies the surface abundances. We show that when initial, bulk metallicity is equated with current, surface metallicity in isochrone age analysis the resulting stellar ages can be systematically over-estimated by up to 20\%. The change of surface abundances with evolutionary phase also complicates chemical tagging, the concept that dispersed star clusters can be identified through unique, high-dimensional chemical signatures. Stars from the same cluster, but in different evolutionary phases, will show different surface abundances. We speculate that calibration of stellar models may allow us to estimate not only stellar ages but also $initial$ abundances for individual stars. In the meantime, analyzing the chemical properties of stars in similar evolutionary phases is essential to minimize the effects of atomic diffusion in the context of chemical tagging.
\end{abstract}


\section{Introduction}
\label{sec:intro}
Atomic diffusion is a generic term used to describe transport processes that operate most effectively in radiative regions in stars. The processes are driven by gradients, including pressure (gravitational settling), temperature (thermal diffusion), and concentration (chemical diffusion). One can find comparisons of the magnitudes of these 3 processes in Figures 3 and 4 of \citet{thoul1994}. The effect of atomic diffusion is greatly reduced in convective regions because the convection timescale is (typically) orders of magnitude shorter than the diffusion timescale. Radiative acceleration modifies atomic diffusion, acting differently on individual species depending in detail on the thermal properties of the plasma and the opacity of each species relative to the total opacity. Including radiative acceleration in the atomic diffusion calculation is computationally intensive \citep{richer1998} and has not yet become a standard ingredient in stellar evolution models, even though atomic diffusion has in recent years. The redistribution of elements during stellar evolution by atomic diffusion may be modified by other mixing process, with convection being the most prominent. While it is beyond the scope of this paper to discuss all mixing processes in stars, we list a few important effects that have been proposed specifically to counteract atomic diffusion in some way: meridional circulation \citep{eddington1929}, mass loss via slow stellar wind \citep{vauclair1995,vick2013}, hydrodynamic instability caused by diffusion itself \citep{deal2016}, as well as a class of `turbulent' mixing schemes \citep{proffitt1991,richer2000} that have been used effectively in different scenarios.

Atomic diffusion has been theorized to operate in stars for more than a century \citep{chapman1917a,chapman1917b}. Decades later, detailed studies arose of its effects on solar models \citep{aller1960,noerdlinger1977} and in other stars \citep[e.g., ][]{michaud1970,montmerle1976,fontaine1979,noerdlinger1980}; see \citet{vauclair1982} for a review of the theory and works completed up to the early 1980s. The importance of atomic diffusion in the calculation of stellar evolution models led to a systematic reduction in age estimates from isochrone fitting to star clusters by $\sim10\%$ compared to models without atomic diffusion \citep[][]{vandenberg2002}.

The use of stellar isochrones to derive the ages of individual stars based on their spectroscopic parameters is widely used \citep{soderblom2010} with a classic example being the estimation of ages for 189 dwarf stars by \citet{edvardsson1993}. Later programs, such as the Geneva Copenhagen Survey  \citep{nordstrom2004}, expanded the numbers of stars into the tens of thousands. In the modern era, surveys like RAVE \citep{steinmetz2006}, SDSS/SEGUE \citep{Yanny2009}, LAMOST \citep{LAMOST}, APOGEE \citep{holtzman2015}, $Gaia$-ESO \citep{gaia-eso}, and GALAH \citep{desilva2015} are yielding hundreds of thousands of stars. These efforts will only improve with the release $Gaia$ parallaxes over the next few years \citep{lindegren2016}. 

In principal, ages may be derived for all of the stars in these surveys. However, given finite measurement uncertainties on global stellar parameters obtainable from spectroscopy, only stars near the main sequence turnoff (MSTO) or the subgiant branch can be age-dated with any reasonable degree of precision. Asteroseismology provides useful information about stellar interiors via the power spectrum of pulsations in dwarfs and red giants \citep{chaplin2013} that has led to novel techniques for estimating stellar masses, and therefore ages, for red giants \citep{martig2016,ness2016}. As a result, although age estimates of red giants cannot be arrived at based on isochrone fitting to spectroscopic parameters alone, the combination of spectroscopic and asteroseismic data opens up a realm of new possibilities.

Stars with measured abundances $and$ ages are extremely valuable quantities in piecing together the formation of different components of the Galaxy. The ability to associate individual stars with their birth environment, i.e., chemical tagging \citep{freeman2002}, is a science driver for the large surveys now underway. Folding in stellar ages, so that these birth clusters can be placed in time as well, adds another dimension to the `detailed physical understanding of the sequence of events which led to the Milky Way' \citep{freeman2002}. Stellar age estimates provide a more direct `clock' than certain abundance ratios (e.g., [Fe/H]) since different populations of Galactic stars have different origins and, thus, different chemical enrichment timescales. 

The main goal of the paper is to emphasize the distinction between the current, surface composition---which can be measured spectroscopically---and the initial, bulk composition---the gas out of which the star formed. In a real star the current, surface composition begins from the initial, bulk composition and is then modified over the star's lifetime by atomic diffusion, nucleosynthesis and mixing processes, and/or accretion. This paper is focused on the role of atomic diffusion, but the interplay between diffusion and other types of mixing is always present. In stellar models the initial, bulk composition is a model $parameter$ (constant in time) while the current, surface composition is a model $prediction$ (variable in time). The two should not be considered identical or interchangeable. If they are, then errors of up to $\sim20\%$ may be introduced in stellar ages (see $\S$ \ref{sec:ages}).

The stellar evolution models used in this paper are based on MESA Isochrones and Stellar Tracks \citep[MIST;][]{dotter2016,choi2016}. The MIST models provide surface abundances for 19 isotopes of 17 chemical elements from H to Fe. These are predictions: they permit new ways of interpreting results from spectroscopic surveys and the opportunity to improve constraints on stellar physics.

\section{Stellar models with atomic diffusion and extra mixing processes}

\subsection{Stellar evolution tracks and isochrones}
\label{sec:MESA}
The stellar evolution code used in this study is Modules for Experiments in Stellar Astrophysics\footnote{\url{http://mesa.sourceforge.net}} \citep[\texttt{MESA};][]{paxton2011,paxton2013,paxton2015} in essentially the same configuration as MIST \citep{choi2016}. The MIST models include atomic diffusion based on the \citet{thoul1994} formalism, exponential-decay overshoot mixing \citep{freytag1996} both below the surface convection zone (if present) and above the convective core (if present), and adopt the \citet{asplund2009} solar abundances. There are 2 significant additions with respect to MIST that we outline here. Both are directly related to atomic diffusion.

First, we have computed stellar evolution tracks both with and without radiative acceleration ($\grad$) in order to gauge its influence. When $\grad$ is enabled, \texttt{MESA} uses monochromatic opacities from the Opacity Project \citep[OP; ][]{seaton2005,seaton2007} to calculate on-the-fly Rosseland mean opacities and $\grad$ for individual species \citep{hu2011,paxton2013}. Each element is treated separately in terms of ionization, monochromatic opacity, and $\grad$ except for Ti. There is no OP data for Ti and, thus, in models with $\grad$ we have combined Ca and Ti into a single class.
Radiative acceleration plays an important role in the diffusion of individual species \citep{richer1998,turcotte1998}.

Second, observations of stars in the metal-poor globular cluster NGC\,6397, which has [Fe/H]=$-2$ \citep{carretta2009} and an age of $\sim13$ Gyr \citep{mf2009,dotter2010,vandenberg2013}, indicate that stellar models with diffusion and $\grad$ still require some additional mixing in the surface layers \citep{korn2007, nordlander2012}. Similar results have been obtained for stars in NGC\,6752 \citep{gruyters2013}.

To inhibit atomic diffusion near the stellar surface \citet{choi2016} used the `radiative diffusivity' of \citet{morel2002}. However, the validity of this treatment has been called into question \citep{alecian2005} and, in practice, it has very little effect in stars like the sun (spectral types G and later). To improve upone this we have incorporated the density-dependent turbulent diffusion coefficient $D_T$ \citep[][eq.\ 5]{proffitt1991} and later modified by \citet{vandenberg2012} to reflect the differing size of the surface convection zone in stars with different initial metallicities.
\begin{equation}
D_{T} = D_0 \left( \frac{\rho_{CZ}}{\rho} \right)^{3} \left( \frac{M_{CZ}}{M_{*}}\right)^{-3/2}
\end{equation}
Where $D_0$ is a constant, $\rho$ is the density, $\rho_{CZ}$ is the density at the base of the surface convection zone, $M_{CZ}$ is the mass of the surface convection zone, and $M_*$ is the mass of the star. We set $D_o = 1~\mathrm{cm^2~s^{-1}}$ in order to best match the observed  $\sim0.2$ dex depletion of metals by \citeauthor{korn2007} in NGC\,6397.

We checked that the calibrated solar model, discussed in detail by \citet[][$\S$4.1]{choi2016}, was still viable with the inclusion of $D_T$ described in the preceding paragraph. For a properly calibrated solar model we require a match to 1 part in $\sim1000$ for the radius and luminosity at the solar age. Solar parameters are taken from the 2015 IAU Resolution B3 \citep{IAU2015B3}. The solar model reproduces the solar surface value $Z=0.0134$ in agreement with \citeauthor{asplund2009} but falls too low in the envelope He abundance, which is consistent with similar models adopting the \citeauthor{asplund2009} solar abundances \citep[see][]{serenelli09}. The solar-calibrated mixing length parameter, $\amlt=1.82$, is unchanged from \citep{choi2016}; the envelope overshoot parameter, $f_{ov,env}$, decreased from 0.0174 to 0.0162. Since $D_T$ and $f_{ov,env}$ operate in a complimentary fashion--both extend mixing below the Schwarzschild boundary--it follows that the inclusion of $D_T$ is compensated by a reduction of $f_{ov,env}$. 

We computed new stellar evolution tracks for $\mathrm{[Fe/H]_{init}}=-2$, $-1.5$, $-1$, $-0.5$, and 0 and $\minit$= 0.5 - $1.5\msun$, except for [Fe/H]=0 which we have extended down to $0.1\msun$. This is a small subset of the masses and, hence, ages covered by MIST but also the most relevant to Galactic Archeology and chemical tagging. All models discussed in the following sections were computed with the same values of $\amlt$, $D_T$, and $f_{ov,env}$. Stellar models with initial masses greater than $1\msun$ have convective core overshoot mixing implemented using the same exponential-decay model as for the envelope with $f_{ov,core}=0.016$ based on calibration to the open cluster M\,67 \citep[see $\S$3.6.2][]{choi2016}. Isochrones were constructed from the tracks as described by \citet{dotter2016}.

\subsection{Model behavior}
\label{sec:behavior}

\begin{figure*}
\centering
\includegraphics[width=0.9\textwidth]{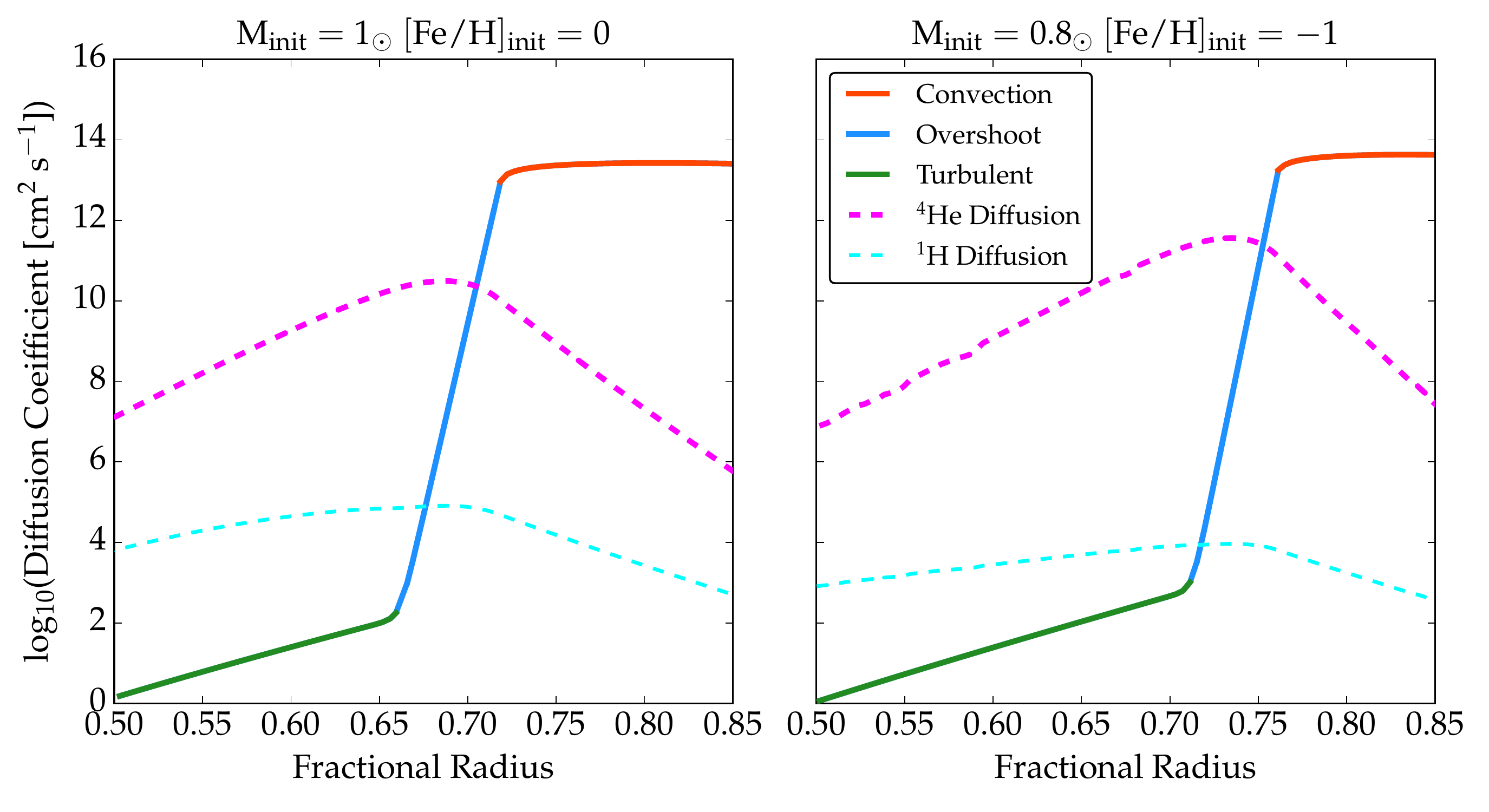} 
\caption{Diffusion coefficient profiles in a solar-type evolutionary track ($\minit=1\msun$, $\mathrm{[Fe/H]_{init}}=0$, $\teff=5832$ K, $\log(g)=4.21$) on the left and a metal-poor evolutionary track ($\minit=0.8\msun$, $\mathrm{[Fe/H]_{init}}=-1$, $\teff=6188$ K, $\log(g)=4.04$) on the right.  The model profiles were taken near the end of core H-burning when $\mathrm{X_H=0.1}$ at the center. Solid lines show the dominant contribution to the diffusion coefficient from standard mixing-length theory convection (red), exponential-decay overshoot mixing (blue), and $D_T$ turbulent mixing (green). Dashed lines show in the same units the results from atomic diffusion for H (cyan) and He (pink).}
\label{fig:coeff}
\end{figure*}

The inclusion of $D_T$ in the stellar evolution tracks has the effect of extending convective boundary mixing below the overshoot region and with a shallower slope. This is demonstrated in Figure \ref{fig:coeff} in which we plot the diffusion coefficient\footnote{Diffusion coefficients, or diffusivities, in the sense of a diffusion equation that is used to solve the mixing of species in \texttt{MESA}.} profiles for two models on the main sequence. Figure \ref{fig:coeff} shows a $\minit=1\msun$ model with $\mathrm{[Fe/H]_{init}}=0$ on the left and a $\minit=0.8\msun$ model with $\mathrm{[Fe/H]_{init}}=-1$ on the right. In both cases, the profiles are shown for models near the end of core H-burning when the central H mass fraction $\mathrm{X_H=0.1}$. The solid line shows the dominant contribution to mixing while the dashed lines show the distinct results of atomic diffusion on H and He. One can see that mixing is extended below the base set by MLT convection with a convective boundary mixing layer, labelled `overshoot', that falls off exponentially \citep{freytag1996}. The overshoot region is extended further, and with a shallower slope, by turbulent diffusion ($D_T$). The net effect of these mixing processes acting below the surface convection zone is a reduced depletion of He and heavier elements from the surface convection zone by atomic diffusion. 

\begin{figure}
  \centering
  \includegraphics[width=0.45\textwidth]{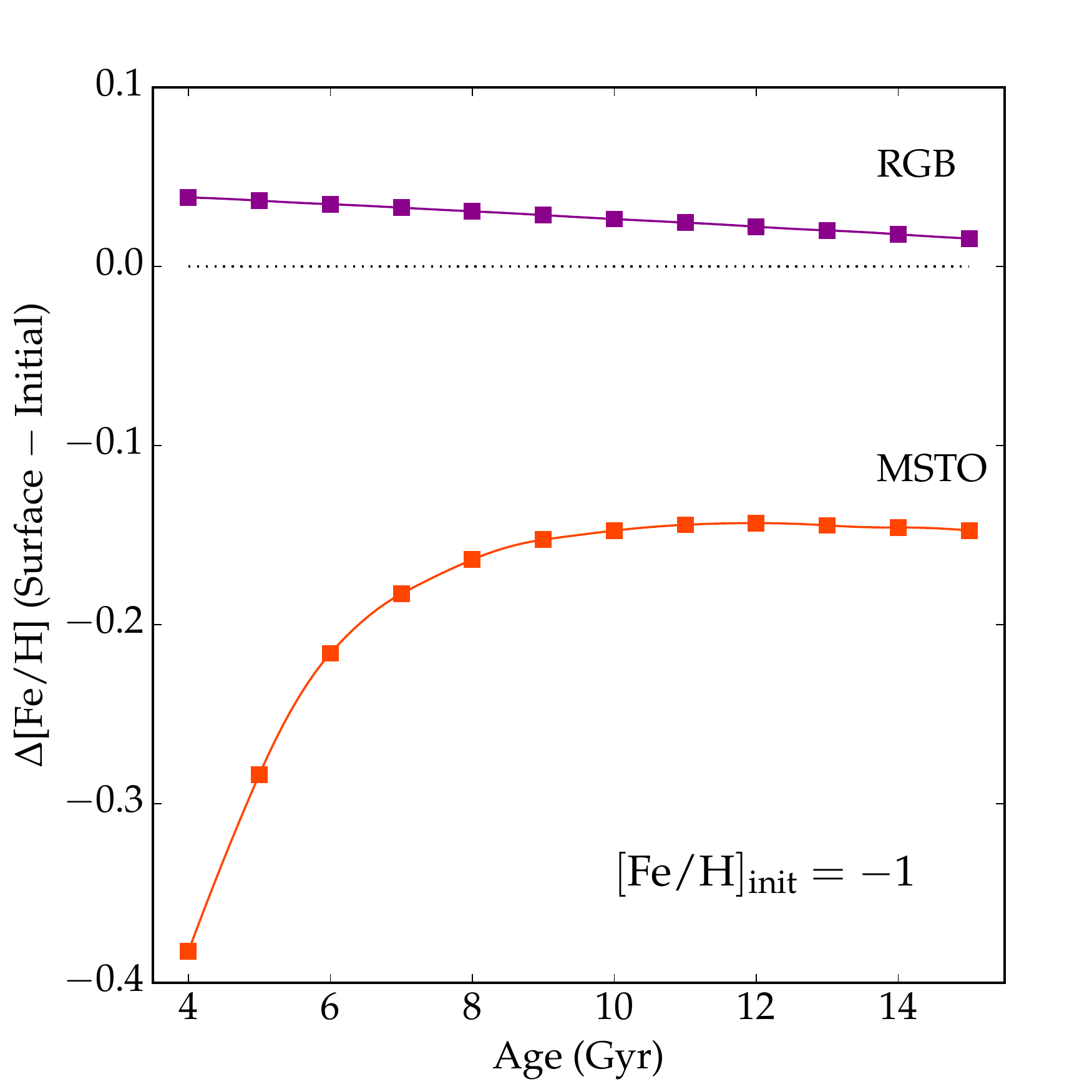}
  \caption{This figure shows the behavior of surface abundance evolution drawn from isochrones with initial $\mathrm{[Fe/H]_{init}=-1}$. Shown are the changes in surface [Fe/H] from the initial value to the MSTO (in orange) and RGB (in purple) as a function of age. The models include atomic diffusion with turbulent mixing but without $\grad$.}
  \label{fig:delta_age}
\end{figure}

\begin{figure}
  \centering
  \includegraphics[width=0.45\textwidth]{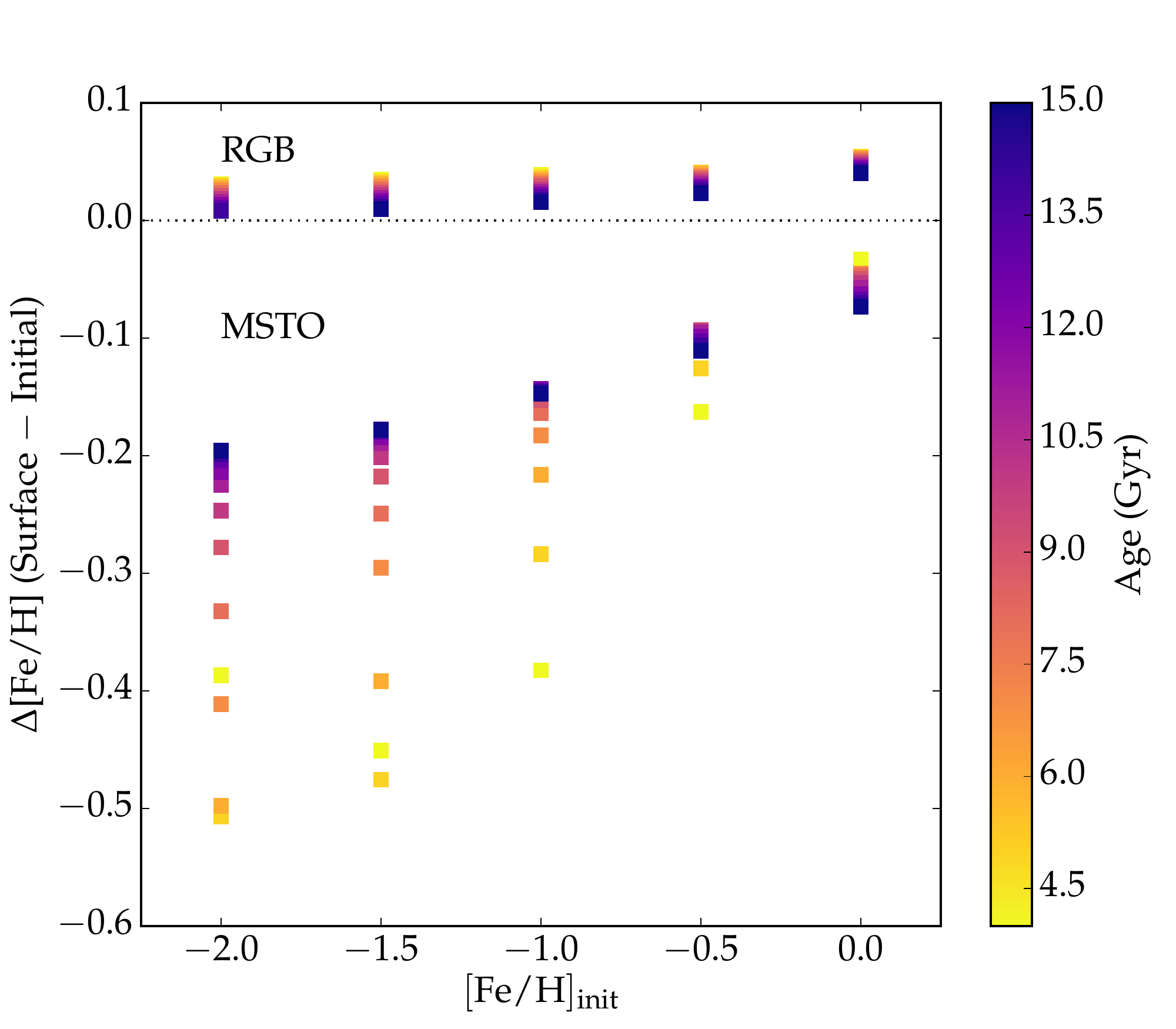}
  \caption{Expanding upon Figure \ref{fig:delta_age} to show the surface [Fe/H] evolution for different ages (reflected in the color scale) and initial metallicities.}
  \label{fig:delta_feh}
\end{figure}

We move now to demonstrations of how the surface [Fe/H] changes as a function of age and $\mathrm{[Fe/H]_{init}}$ at different locations in the H-R diagram.
Figure \ref{fig:delta_age} shows how the surface [Fe/H] value evolves with age from 4 to 15 Gyr for a population with $\mathrm{[Fe/H]_{init}}=-1$. The lower data points, shown in red and labelled `MSTO', show how the surface value decreases from the initial due to atomic diffusion just below the MSTO. The MSTO masses range from $0.75\msun$ at 14 Gyr to $1.05\msun$ at 4 Gyr. The depletion is largest at younger ages when the surface convection zone is smallest and, thus, diffusion is most efficient. In cases where the stars near the MSTO maintain a convective core the location of maximum depletion is near the hottest point on the isochrone \emph{below the convective hook}. The definition of the MSTO used in the models (central H exhaustion) may differ from the observational definition (e.g., the hottest point in the CMD) and therefore lead to a discrepancy. Thus, when comparing models to data is important to choose the point(s) for comparison consistently.
The upper data points, shown in purple and representing stars on the red giant branch (RGB), show how the surface [Fe/H] value returns to near its initial value after the main sequence is over, surface convection deepens, and the near-initial surface composition returns. The difference between the initial and RGB values is $not$ zero, in fact it is slightly positive, because central H-burning has reduced H. When this material is brought back up to the surface by convective mixing (i.e., the first dredge-up), the result is a slight increase in [Fe/H] over the initial value. Along the RGB the surface [Fe/H] varies by less than 0.01 dex. The difference between the MSTO and the RGB surface [Fe/H] values is, thus, a modest overestimate of the atomic diffusion effect compared to the initial value.

Figure \ref{fig:delta_feh} expands upon the models shown in Figure \ref{fig:delta_age} by including all other initial metallicities computed for this paper.  The range of ages is now represented by the color scale. The increase in [Fe/H] from the initial value to the RGB value is never more than about 0.05 dex, while the decrease in [Fe/H] at the MSTO is larger and has significant dependence on both the age and $\mathrm{[Fe/H]_{init}}=0$. At the oldest ages and lowest metallicities, however, the decrease at the MSTO is $\sim0.2$ dex, consistent with NGC\,6397 \citep{korn2007}. The mass of the surface convection on the main sequence tends to increase with metallicity at a given mass and so the MSTO depletion increases substantially with decreasing $\mathrm{[Fe/H]_{init}}=0$.

\begin{figure}
  \centering
  \includegraphics[width=0.45\textwidth]{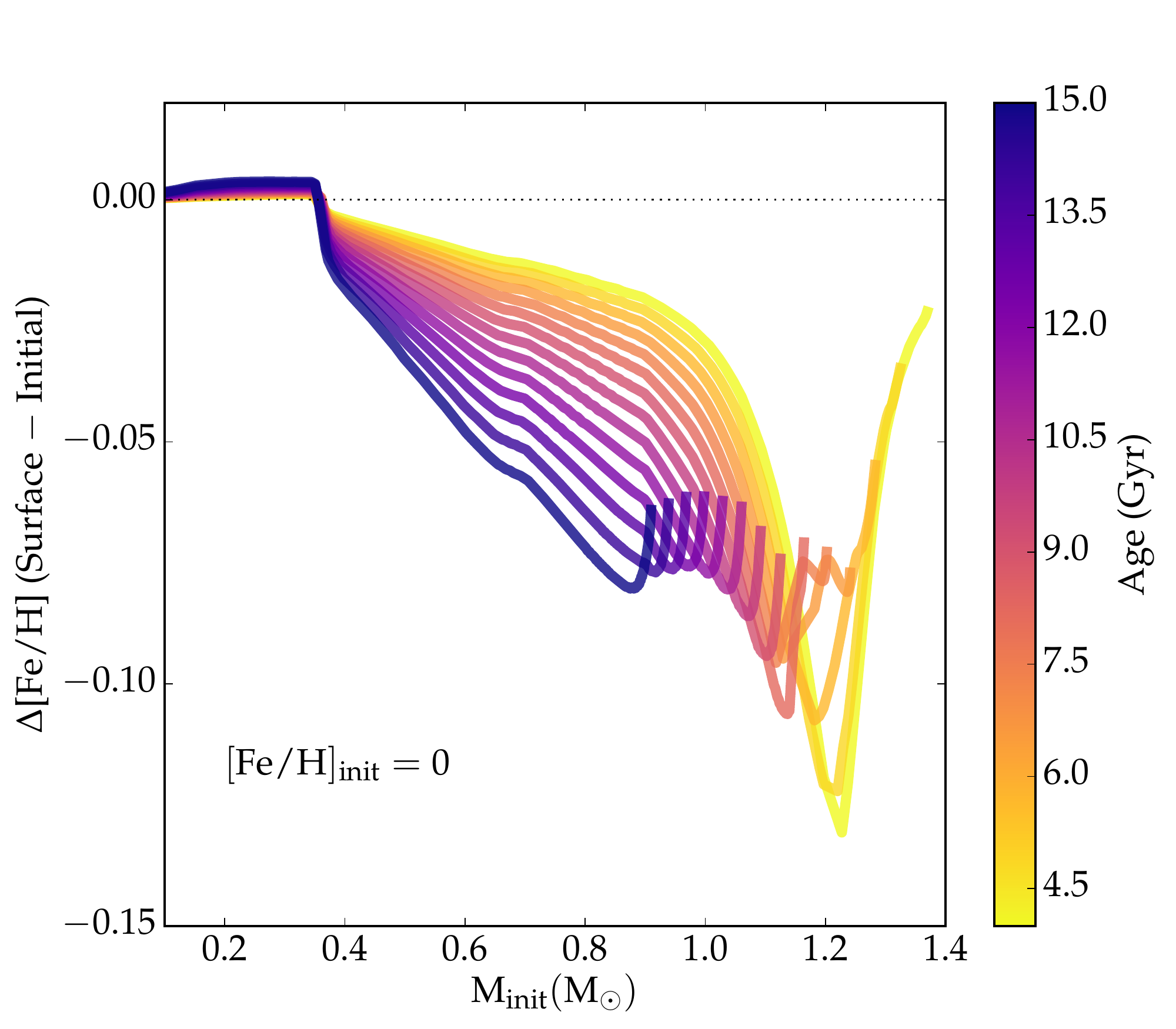}
  \caption{The changing surface [Fe/H] along the main sequence for a range of ages. The sharp transition at $\sim0.35 \msun$ marks the boundary between fully-convective stars at lower masses and stars with radiative cores at higher masses. }
  \label{fig:MS}
\end{figure}

In a coeval, initially-chemically-homogeneous stellar population, like an open cluster, the farther down the main sequence one looks, the closer the surface abundances will approach the intial abundances of a cluster, due to the deepening of the surface convection zone as $\minit$ decreases as demonstrated in Figure \ref{fig:MS}. This raises the question: Is there a point along the main sequence below which the current, surface abundances match the initial abundances? In the following discussion we assume that all stars have arrived on the main sequence. 
For stars with $\minit < 1.5 \msun$ surface convection deepens as one moves along the main sequence to lower $\minit$, meaning that atomic diffusion will be less effective at modifying surface abundances. This can be seen in Figure \ref{fig:MS} as $\Delta$[Fe/H] decreases with $\minit$. There is a point along the main sequence, at $\minit \sim 0.35 \msun$, below which stars remain fully-convective on the main sequence. For a fully-convective star atomic diffusion is overwhelmed by convective mixing. Furthermore, the effects of nucleosynthesis ought to appear at the surface as a slight increase because a fully-convective star is fully mixed, see Figure \ref{fig:MS}. We expect the surface depletion of metals, as measured by $\Delta$[Fe/H] in Figure \ref{fig:MS} to: (i) reach its maximum value near the MSTO (just below if the stars have convective cores); (ii) decrease as a function of stellar mass down to the fully-convective transition at $0.35 \msun$; and (iii) abruptly become slightly \emph{positive} because H has been depleted by nucleosynthesis in the core. The fully-convective stars are intrinsically faint and so the transition will be difficult to detect in practice. However, some of the larger patterns, such as the trend of [Fe/H] at the MSTO with age, should be detectable.

\section{Implications}
\label{sec:implications}

\subsection{Individual stellar ages from spectroscopic parameters}
\label{sec:ages}

Suppose that we wish to estimate the age of a star based on its spectroscopic parameters. We are given $\teff$, $\logg$, [Fe/H], and associated uncertainties. We compare these with a set of stellar isochrones in order to discern what range of stellar mass and age are consistent with the models given the data. The age analysis is carried out using one set of stellar evolution models with atomic diffusion, as described in $\S$\ref{sec:MESA}, and two different assumptions. In the first case, only the initial composition is specified in the models. We have no idea how the surface abundances evolve as a function of $\minit$ and time. Thus, we ignore any change in surface abundances \emph{along the isochrones}. Instead, we match the initial abundances in the model to the observed surface abundances in each star. We shall refer to this as the `constant-metallicity' assumption. In the second case, the surface abundances are tabulated along the isochrones and we are able to treat these as variables. We shall call this the `variable-metallicity' assumption. In the context of atomic diffusion, the problem with the constant-metallicity assumption is that the interior of the star in question can actually be more metal-rich than the current, surface abundances indicate. The variable-metallicity assumption naturally accounts for this effect, albeit in a model-dependent way.

Ages are estimated with \texttt{MINESweeper}, a Bayesian approach to estimating stellar parameters for a single star from stellar evolution models. A full description of \texttt{MINESweeper} is given by Cargile et al.\ (in prep). Here we provide a brief summary of the method. \texttt{MINESweeper} uses nested importance sampling to determine posterior probability distributions for physical properties inferred from stellar evolution models. It uses a modified version of the \texttt{nestle.py} code\footnote{\url{http://kbarbary.github.io/nestle/}} to perform multi-nested ellipsoid sampling based on the algorithm described in \citet{feroz2009}. Multi-nested sampling is well-suited for this problem due to its ability to efficiently sample multi-modal likelihood surfaces, as is typically the case when modeling stars with stellar isochrones. \texttt{MINESweeper} uses the most recent release of the MIST stellar evolution models \citep{choi2016}, and an optimized interpolation schema based on the recommendations of \citet{dotter2016}. \texttt{MINESweeper} uses photometry, spectroscopic parameters and abundances, parallax information, etc.\ to calculate the likelihood probabilities, and handles a wide range of prior probability distributions. The \texttt{MINESweeper} inference results in posterior probability distributions for the fundamental MIST model parameters (i.e., the parameters on which the model grid is built): equivalent evolutionary points (EEPs), initial metallicity, and stellar age. Using these inferred posterior distributions, \texttt{MINESweeper} then generates posterior distributions for all other quantities predicted by the MIST models, e.g., mass, radius, $\teff$, luminosity, surface abundances, etc.

We take as a case study the spectroscopic survey of the Galactic disk by \citet[][hereafter B14]{bensby2014}. B14 derived spectroscopic abundances and ages for 714 stars, primarily dwarfs and subgiants of spectral type F or G, with $5 > \logg > 2.5$, in the Solar neighborhood. The method of age estimation employed by B14 is described in detail by \citet{melendez2012} and uses the $\Ysq$ isochrones \citep[][]{demarque2004}, which include the effects of He diffusion but not of heavier elements.

The age analysis employed by B14 relies on comparing the spectroscopic parameters $\teff$, log(g), and [Fe/H] with those of the $\Ysq$ isochrones. The $\Ysq$ isochrones do not provide current, surface [Fe/H] along the isochrones so the age analysis uses instead the associated $\mathrm{[Fe/H]_{init}}$ for all ages and all points along the isochrones. As described by \citeauthor{melendez2012}, in order to reconcile the isochrone composition with the current solar values, the authors subtract 0.04 dex from the tabulated [Fe/H] values in the $\Ysq$ isochrone files. This is, in effect, a correction for the diffusion of elements from the solar surface over $\sim$4.5 Gyr \citep{turcotte2002,asplund2009}. Although \citeauthor{melendez2012} do not state this explicitly, nor is it clear that they interpret it as such, they have implemented a crude correction for the difference between current, surface [Fe/H] and the initial, bulk value. However, as shown in Figures \ref{fig:delta_age} and \ref{fig:delta_feh}, the difference between the current, surface [Fe/H] and the initial, bulk [Fe/H] is not constant over either a single isochrone or among different isochrones (see Figures \ref{fig:delta_age} and \ref{fig:delta_feh}). This is not a criticism of the age analysis technique described by \citeauthor{melendez2012}; their equation 7 $would$ have properly accounted for changing surface abundances along the isochrones if that information had been provided in the isochrones themselves. 

\begin{figure}
\centering
\includegraphics[width=0.45\textwidth]{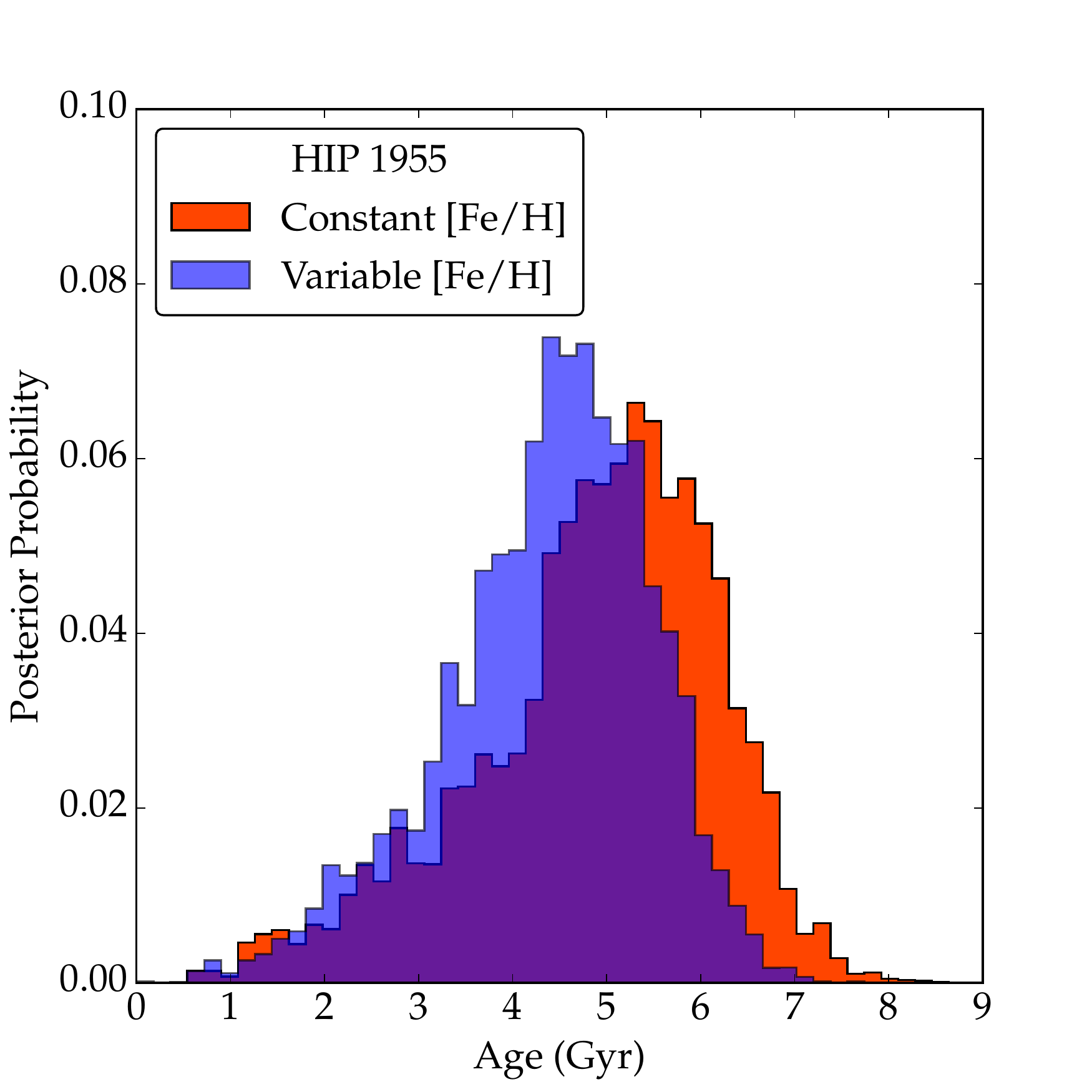} 
\caption{Posterior probability distributions from two age analyses of a single star, HIP 1955, using spectroscopic parameters from \citet{bensby2014}. The blue histogram shows the case in which the isochrones' initial [Fe/H] value is compared to the spectroscopic value. The red histogram shows the case in which the isochrones' surface [Fe/H] value is compared to the spectroscopic value. The latter is the approach we advocate for in this paper; it results in a younger age.}
\label{fig:starpar}
\end{figure}

To illustrate the difference in stellar ages between isochrone analyses that use constant or variable surface metallicities, we have derived ages for the B14 catalog using both the constant- and variable-metallicity assumptions. To do so we use $\teff$, $\logg$, and [Fe/H] from the B14 catalog.
In Figure \ref{fig:starpar} we show the age distributions for HIP 1955 based on the spectroscopic parameters ([Fe/H]=$-0.01$, $\teff=6024$, and $\logg=4.31$) provided by B14. The red histogram shows ages obtained using the variable-metallicity approach. The blue histogram shows ages obtained using the constant-metallicity approach. The variable-metallicity approach results in a weighted-mean age of 4.37 Gyr while the variable-metallicity approach yields a weighted-mean of 4.89 Gyr. The result is a reduction in the age of $\sim0.5$ Gyr, or about 10\%. Weighted medians are $\sim0.1$ Gyr younger than the weighted means for this star in both cases but the difference remains about 0.5 Gyr.

\begin{figure*}
\centering
\includegraphics[width=0.9\textwidth]{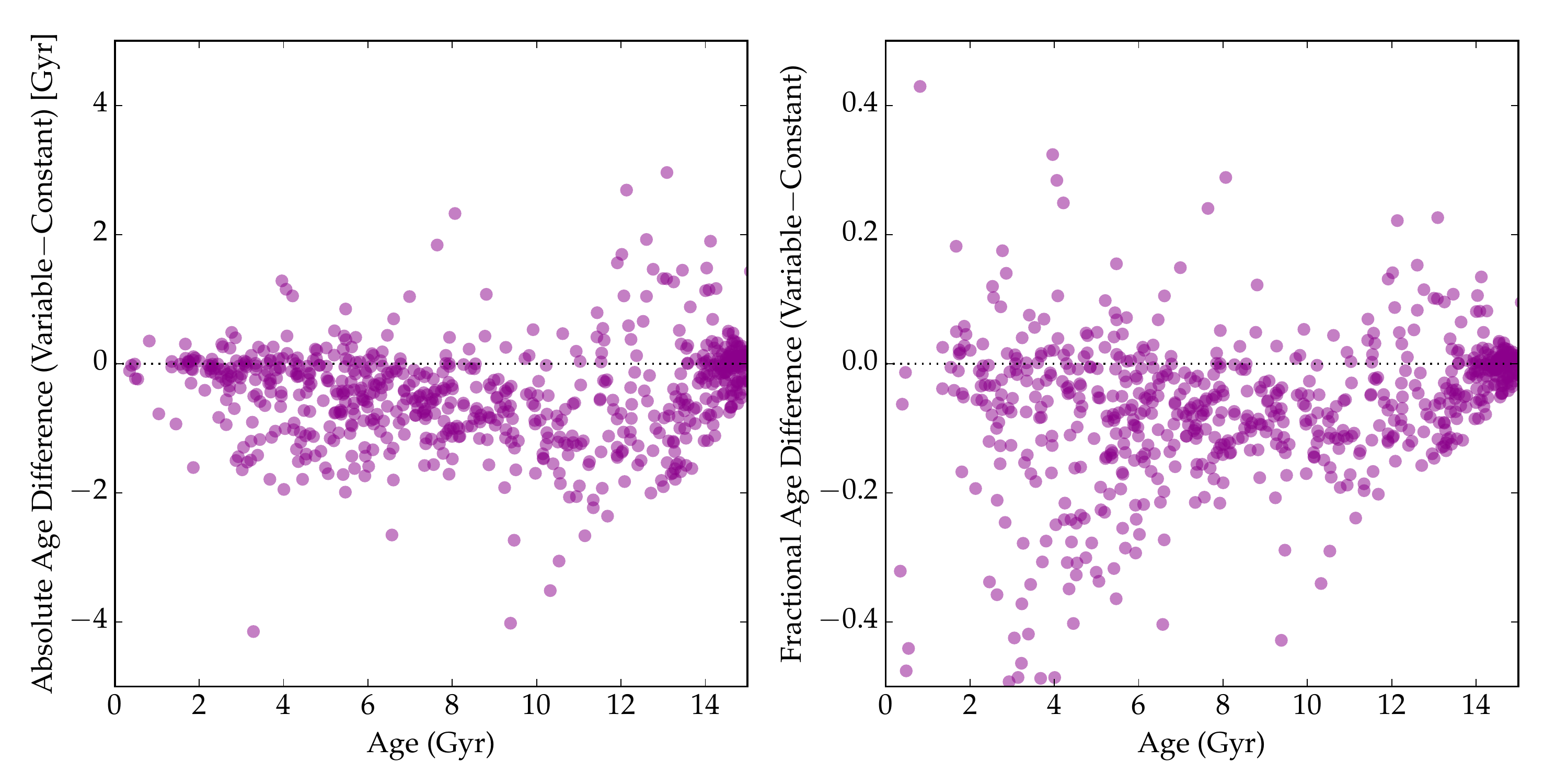} 
\caption{Difference in ages determined using the variable- and constant-metallicity assumptions. The difference is most often negative. The left panel shows age differences in an absolute sense and the right panel shows them in a fractional sense. Note that the majority of points are negative, meaning that the constant metallicity assumption systematically increases the age.}
\label{fig:delta_age_bensby}
\end{figure*}

In Figure \ref{fig:delta_age_bensby} we show the age difference between the two assumptions for the full B14 catalog in both absolute (left) and fractional (right) terms. Both panels show that the ages based on the variable-metallicity approach lead to younger ages than those derived by the constant-metallicity approach.  This systematic effect in isochrone-based age determinations of individual stars has largely been overlooked until now.

\begin{figure*}
\centering
\includegraphics[width=0.9\textwidth]{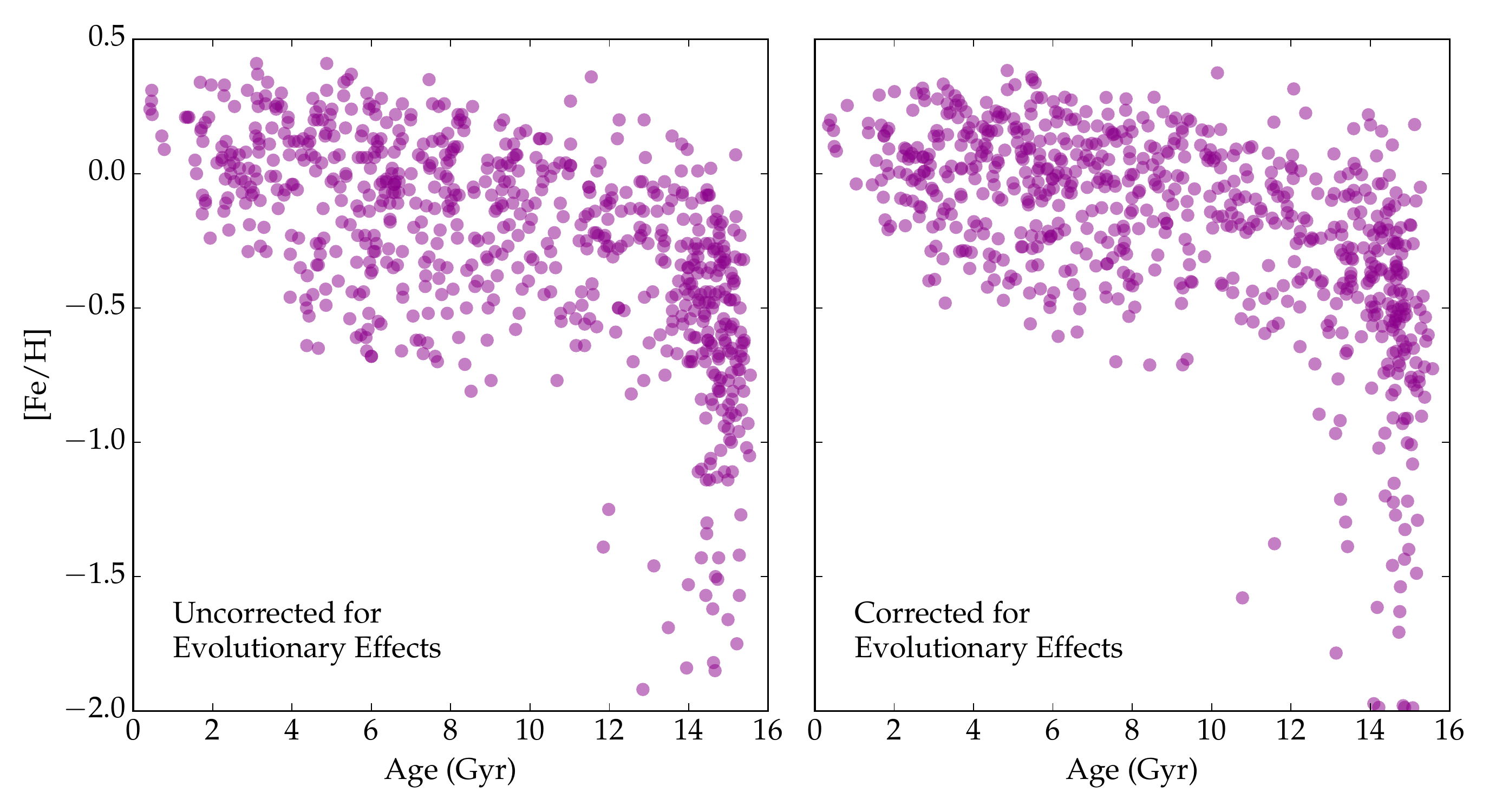} 
\caption{The AMR of \citet{bensby2014} stars re-derived in this paper. On the left are shown surface [Fe/H] values vs.\ ages derived using the constant-metallicity assumption \citeauthor[as done by ][]{bensby2014}. On the right are shown (inferred) $initial$ metallicities and ages derived using the variable-metallicity approach.}
\label{fig:amr}
\end{figure*}

A corollary is that the initial, bulk metallicity of a star---not the current, surface metallicity---should be the quantity of interest in age-metallicity relations (AMRs) and Galactic chemical evolution (GCE). The reason is that the initial, bulk metallicity is reflective of the conditions out of which the star formed $and$ is a stellar parameter, not a variable quantity. The result will be an age-metallicity relation that is shifted to younger ages $and$ higher (initial, bulk) metallicities. Figure \ref{fig:amr} shows the B14 sample in the standard AMR that plots current, surface metallicities and constant-metallicity-based ages on the left.  On the right is the shown the corrected version of the AMR that plots the inferred initial, bulk metallicity of the star against the variable-metallicity-based age. Note that only stars with [Fe/H] $> -2$ are shown, though there are a handful of stars in the B14 catalog with metallicities below $-2$.

Initial, bulk metallicities are inferred from the models and, thus, susceptible to model errors and uncertainties. It will be valuable to provide more detailed tests of the models so that these ``corrections'' may be applied with greater confidence. Open clusters are highly useful in this regard because they come with a high degree of chemical homogeneity \citep[][but see also \citet{liu2016b}]{bovy2016}. Globular clusters have also been studied in this context, including NGC\,6397 \citep{korn2007,lind2009,nordlander2012}, NGC\,6752 \citep{gruyters2013}, and M\,30 \citep{gruyters2016}, despite the prevalance of internal abundance variations \citep{gratton2004}.

\subsection{Solar analogs}
\begin{figure*}
  \centering
  \includegraphics[width=0.9\textwidth]{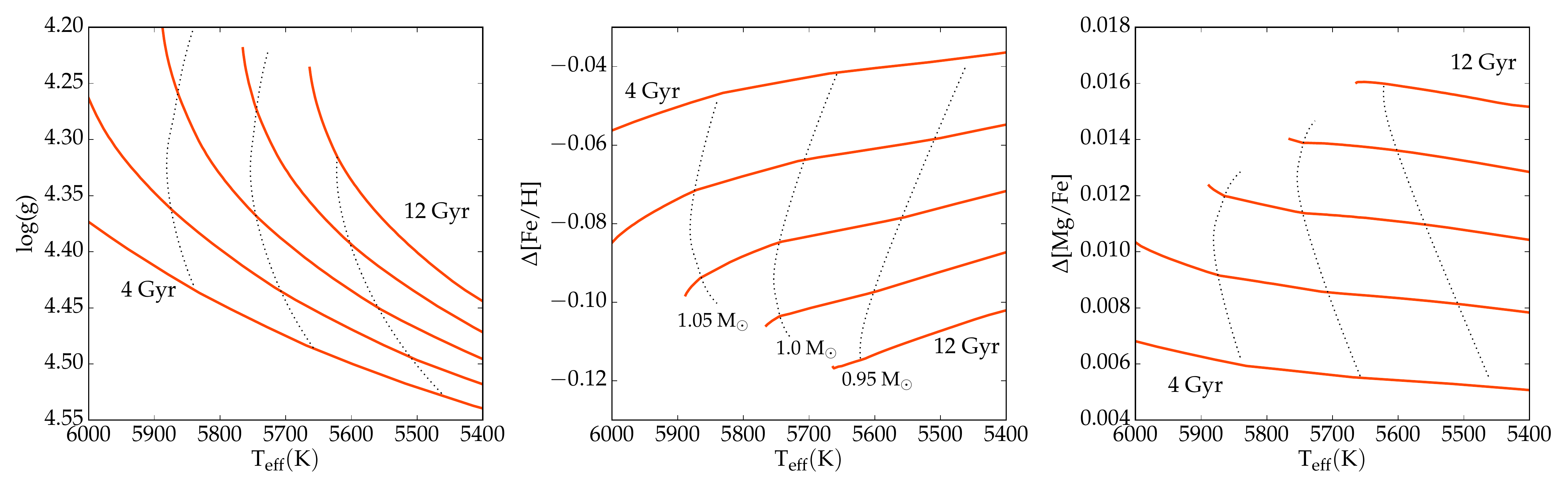}
  \caption{Model behavior for solar analogs spanning a range of ages. The models all have $\mathrm{[Fe/H]_{init} = 0}$. Isochrones with ages of 4, 6, 8, 10, and 12 Gyr are plotted as solid lines while evolutionary tracks for $\minit=0.95, 1.00, 1.05 \msun$ are shown as thin, dotted lines. The left panel shows the range of $\teff$ and $\logg$ covered while the center and right panels show the depletion of [Fe/H] and [Mg/Fe] relative to the initial value in the models.}
  \label{fig:solar}
\end{figure*}

Stars with observable properties similar to those of the Sun are of astrophysical interest for many reasons, including the search for exoplanets and the uniqueness of our solar system \citep{cayrel1996}. Atomic diffusion is an important ingredient in standard solar models \citep{models,asplund2009} and is, by extension, relevant in studies of solar analogs. By solar analogs we mean MS stars within roughly 300K of the solar $\teff$ and $-0.1 <$ [Fe/H] $< +0.1$. Figure \ref{fig:solar} shows stellar evolution models with $\mathrm{[Fe/H]_{init} = 0}$. Isochrones with ages ranging from 4 to 12 Gyr are shown as solid lines while stellar evolution tracks with $\minit=0.95, 1.00, 1.05 \msun$ are shown as dotted lines. The left panel shows the range of $\teff$ and $\logg$ considered in the center and right panels.  The center panel shows the surface evolution of [Fe/H] relative to the initial value. The right panel shows the surface evolution of [Mg/Fe], which is indicative of a lighter metal than Fe. Mg has a lower atomic mass and nuclear charge than Fe, and so it settles marginally more slowly than Fe, resulting in a flat or slightly-positive [Mg/Fe] trend with $\teff$ at fixed age in the right panel of Figure \ref{fig:solar}. This is the opposite of [Fe/H] shown in the middle panel of Figure \ref{fig:solar}.

Stellar evolution models that include atomic diffusion predict a systematic trend in surface [Fe/H] as a function of $\teff$ in the realm of solar analogs. In a coeval, initially-chemically-homogeneous sample this trend should be observable at the current measurement precision of $\sim0.01$ dex \citep[e.g., ][]{melendez2017}. While the [Mg/Fe] trend shown here may be below the current threshold, we recommend pursuing this and other ratios to provide further constrain the models.

\subsection{Chemical Tagging}
\label{sec:tagging}

\begin{figure*}
  \centering
  \includegraphics[width=0.9\textwidth]{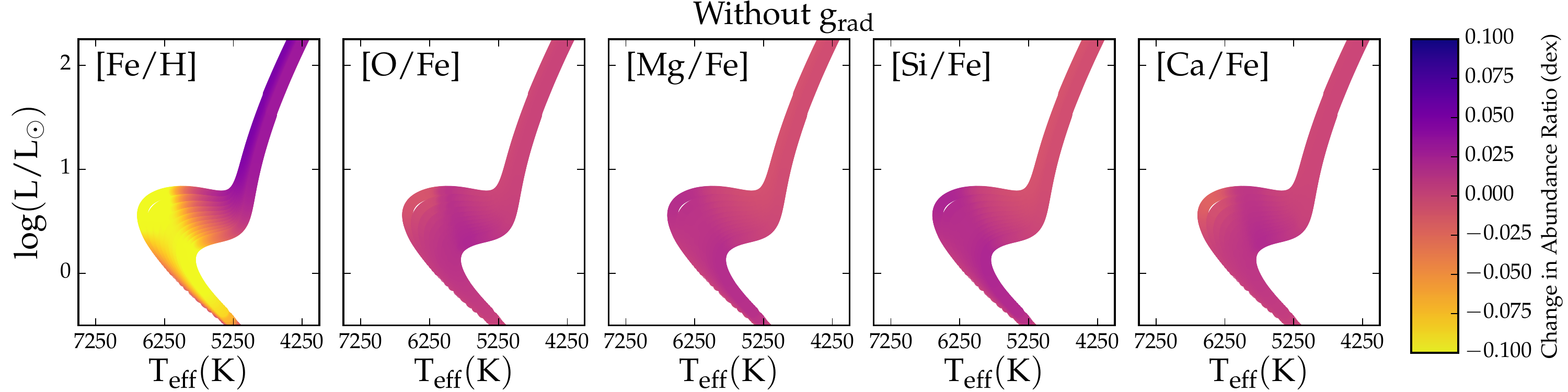}

  \vspace{0.5cm}
  
  \includegraphics[width=0.9\textwidth]{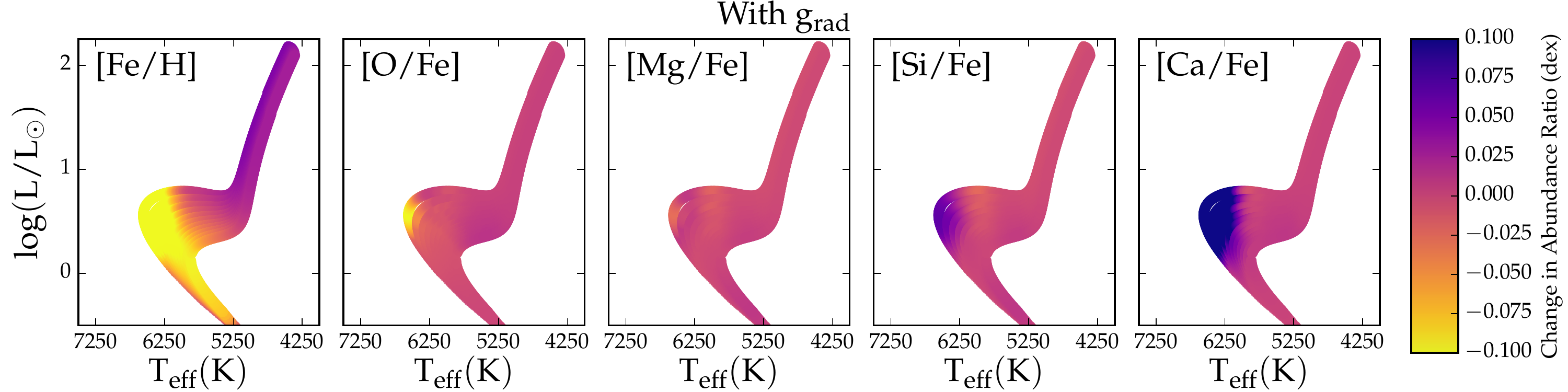}
  \caption{Evolution of surface abundance ratios in the H-R diagram for models with atomic diffusion and turbulent mixing; the top panel shows models without $\grad$ while the bottom row shows models with $\grad$. The models have $\mathrm{[Fe/H]_{init}}=-0.5$ and ages from 5 to 15 Gyr. The leftmost column shows the variation of Fe \emph{with respect to H} while the other columns show variations \emph{with respect to Fe}.}
  \label{fig:tag}
\end{figure*}

Chemical tagging, in which individual stars are associated with their birth cloud via its unique chemical signature, promises a new view of the Galaxy and its assembly over time \citep{freeman2002}. Although chemical tagging as originally defined by \citeauthor{freeman2002}\ has yet to be realized, progress is being made \citep[see ][]{hogg2016,ting2016}. Furthermore, the case in which individual stars are identified as originating in a certain $type$ of object, but not one $specific$ object, has already seen some success \citep{martell2011,martell2016,schiavon2016}.

Chemical tagging should take into consideration the fact that stars born in the same birth cloud, but now in different evolutionary phases (because they have different initial masses), will show signs of atomic diffusion that confuse the sought-after chemical signature. Working in absolute abundance space (or relative to H) will expose this issue whereas abundance ratios, such as [X/Fe], will reduce---but not remove---the effects of atomic diffusion. Studying stars that are in the same evolutionary phase, such as red giants, will also simplify the analysis. The uncertain extent to which $\grad$ in stars of different spectral types plays a role in surface abundances must also be considered.

We show examples of how surface abundance variations manifest across the H-R diagram in Figure \ref{fig:tag}. It shows models with ages from 5 to 15 Gyr and $\mathrm{[Fe/H]_{init}=-0.5}$. The models shown include atomic diffusion and turbulent mixing; the top row doesn not include $\grad$ while the bottom row does. The main difference between the two is that the models with $\grad$ show variations around the MSTO that are absent in the models without $\grad$. This is consistent with the nonlinear behavior of $\grad$ with atomic number \citep[compare with Figure 1 of ][]{korn2007}.

In the context of chemical tagging, in which a unique chemical signature is sought, spectroscopic surveys that target red giants, like APOGEE, have an advantage in that the diffusion effect is minimized in these stars.  In surveys that target stars in all evolutionary phases, like GALAH,  clump-finding algorithms can be set to work on abundance ratios, e.g., [X/Fe], in which diffusion effects are reduced. In the latter case, as demonstrated by Figure \ref{fig:tag}, searching for clumps in absolute abundances ([X/H]) will be more challenging than searching in abundances ratios ([X/Fe]).

A new approach to chemical tagging is to use stellar evolution models to infer the initial, bulk abundances from the (observed) current, surface abundances in individual stars and $then$ look for clumps in abundance space.  This is obviously a(nother) step removed from looking for clumps directly in the data and requires confidence in the models. While ambitious, this approach offers the possibility of correcting for evolutionary effects and may prove fruitful in the future, but only with thoroughly-calibrated models.

Recently, \citet{ness2017} introduced the concept of ``stellar doppelg{\"a}ngers'', i.e., pairs of stars with indistinguishable chemical abundance patterns (within measurement uncertainties) that were not born in the same birth cloud. These stars would register as false positives in chemical tagging. Not only can atomic diffusion cause two coeval stars of different $\minit$ (hence evolutionary phase) born in the same birth cloud to have different surface abundances but it can also create these doppelg{\"a}ngers. Such coincidental pairs can arise between two stars in different evolutionary phases if the effect of atomic diffusion (e.g., $\Delta$[Fe/H]$= -0.05$) is offset by an equal and opposite difference in the initial abundances (e.g., $\Delta$[Fe/H]$\mathrm{_{init}} = +0.05$). This assumes the rest of the abundance pattern scales with Fe, which is not unreasonable at the sub-$0.1$ dex level.

To summarize our argument, chemical tagging studies should take into account our current knowledge of atomic diffusion. At present, perhaps the most sensible approach is to consider only stars in the same evolutionary phase---the tighter the constraints, the better. In the future, we hope that it will be possible to use stellar evolution models to correct for atomic diffusion effects so that stars in different evolutionary phases can be safely studied together.

\section{Summary}
How much can we trust stellar evolution models? All the conclusions in this paper are based on one set of models computed with one stellar evolution code (\texttt{MESA}) with one set of physical assumptions oulined in $\S$\ref{sec:MESA}. While another set of models might differ quantitatively from what we have presented in this paper, the qualitative picture should not change. With the present and future of asteroseismology, the promise of the $Gaia$ mission, and the wealth of stellar parameters and abundances from spectroscopic surveys, stellar models will be subject to renewed scrutiny in the coming years. Critical tests of atomic diffusion and other mixing processes in single stars of similar age and initial composition, but different evolutionary phases, are the most useful in this context. Such studies are, however, few and far between at present.

We highlight the importance of tabulating surface abundances along isochrones. Model surface abundances are necessary when deriving accurate individual stellar ages from spectroscopic parameters and abundances. Ages estimated from isochrones based on the (incorrect) constant-metallicty assumption may be overestimated by as much as 20\% compared to ages derived via the variable-metallicity assumption.
When applied to large samples, this has important implications for studies of the AMR in Galactic field stars and GCE. However, it remains difficult to properly address this issue because many stellar isochrones either (i) ignore atomic diffusion and/or (ii) omit the surface abundance information along the isochrone even though this information is computed by all stellar evolution code.

Chemical tagging is complicated by evolutionary effects in the surface abundances of stars. This problem can be mitigated by either choosing stars in the same evolutionary phase and/or by focusing on abundance ratios (i.e., [X/Fe]) rather than absolute abundances ([X/H]), though we appreciate that some form of metallicity ([Fe/H]) is important. We suggest that, after thorough calibration of stellar models, a new approach to chemical tagging will become possible. The approach uses the stellar evolution models to revert observed surface abundances back to their initial, bulk values. Chemical tagging can then be attempted directly in initial abundance space, thereby avoiding evolutionary effects. We are a long way from implementing this in practice but it presents an intriguing possibility.

\acknowledgments{We thank the anonymous referee for providing a helpful and constructive report.  Special thanks to Bill Paxton and all who have contributed to \texttt{MESA}. AD, CC, and PC received support from NSF grant AST-1313280 and the Packard Foundation. AD and MA were supported by the Australian Research Council under grant FL110100012.}

\end{document}